# Evolutionary constraints on protein fold diversity

Harutyun Sahakyan[a], Eugene V. Koonin[a,*]

a - Computational Biology Branch, Division of Intramural Research, National Library of Medicine, NIH, Bethesda, MD 20894

* - Corresponding author (koonin@ncbi.nlm.nih.gov)

***Abstract***

Arrangements of secondary structures in proteins yield a wide variety of possible topologies, yet the protein folds observed in nature are limited in their diversity. Is the number of protein folds limited by biophysical constraints or by evolutionary contingency? In recent years, the development of new methods for protein structure prediction and comparison, molecular design, and the rapid progress of cryo-electron microscopy revealed new insights into the protein fold space. Here, we present recent findings demonstrating that natural proteins adopt only a small fraction of all possible folds and discuss how evolutionary processes limit the diversity of evolving protein folds.

## 1. Introduction

Although there is an astronomical number of foldable protein sequences, the number of distinct protein architectures – folds - observed in nature is comparatively small. If a protein fold is defined by its core structure and topological arrangement of the structural elements, irrespective of sequence, there are slightly over 1,000 characterized protein folds (Andreeva *et al.*, 2020, Waman *et al.*, 2025, Schaeffer *et al.*, 2026). By contrast, tens of thousands of functional protein families have been discovered through sequencing and comparative genomics, so that many folds accommodate numerous families each (Blum *et al.*, 2025, Paysan-Lafosse *et al.*, 2025). Despite the rapidly growing number of sequenced genomes and metagenomes, the trend is that the number of novel protein folds characterized each year has been declining (Sato & Tomii, 2026, Sykes *et al.*, 2023), suggesting a limited diversity of distinct folds in nature. Even analysis of more than 200 million predicted protein structures, three orders of magnitude more than there are experimentally determined structures in the Protein Data Bank (PDB), yielded fewer than 10,000 distinct folds (Lau *et al.*, 2024).

It has been suggested that the slowing discovery rate of novel protein folds is due to natural evolution having exhausted the possible architectures of small protein domains, and that most of the folds have already been characterized (Zhang *et al.*, 2006). However, the opposing view is that the combinatorial space of potential stable protein folds is much greater than what we observe in nature, but evolution simply has not sampled the great majority of the possible folds. Under this view, evolution proceeds predominantly by tinkering with and repurposing existing protein domains rather than by creating new ones from scratch or by radical rewiring of existing folds. Accordingly, a relatively small core set of folds accounts for the majority of all globular proteins, whereas a long tail of the domain abundance distribution consists of rare protein architectures (Zeldovich *et al.*, 2007, Orengo *et al.*, 1994).

With the development of new protein design methods, it becomes possible to explore the protein fold space more efficiently and to address such questions at the quantitative level (Kuhlman *et al.*, 2003). Minami et al. computationally designed all possible αβ-protein folds with a central β-sheet and surrounding α-helices connecting the β-strands (Minami *et al.*, 2023). Experimental determination of the structures of 8 novel αβ folds containing a central 4-stranded β-sheet with different architectures demonstrated that these folds are indeed stable and can be potentially functional. Other topologies with larger β-sheets encompass multiple novel folds, with an estimate of more than 10,000 novel αβ-folds containing β-sheets with 5-8 strands. Similarly, multiple novel all-α folds were computationally generated, and the structures of 5 of these proteins were solved experimentally (Sakuma *et al.*, 2024). Although these studies were focused on particular topologies, the number of the identified novel domains substantially exceeds the number of cataloged protein folds. Thus, the number of possible stable protein folds should be much larger than the number of observed ones (Fig. 1).

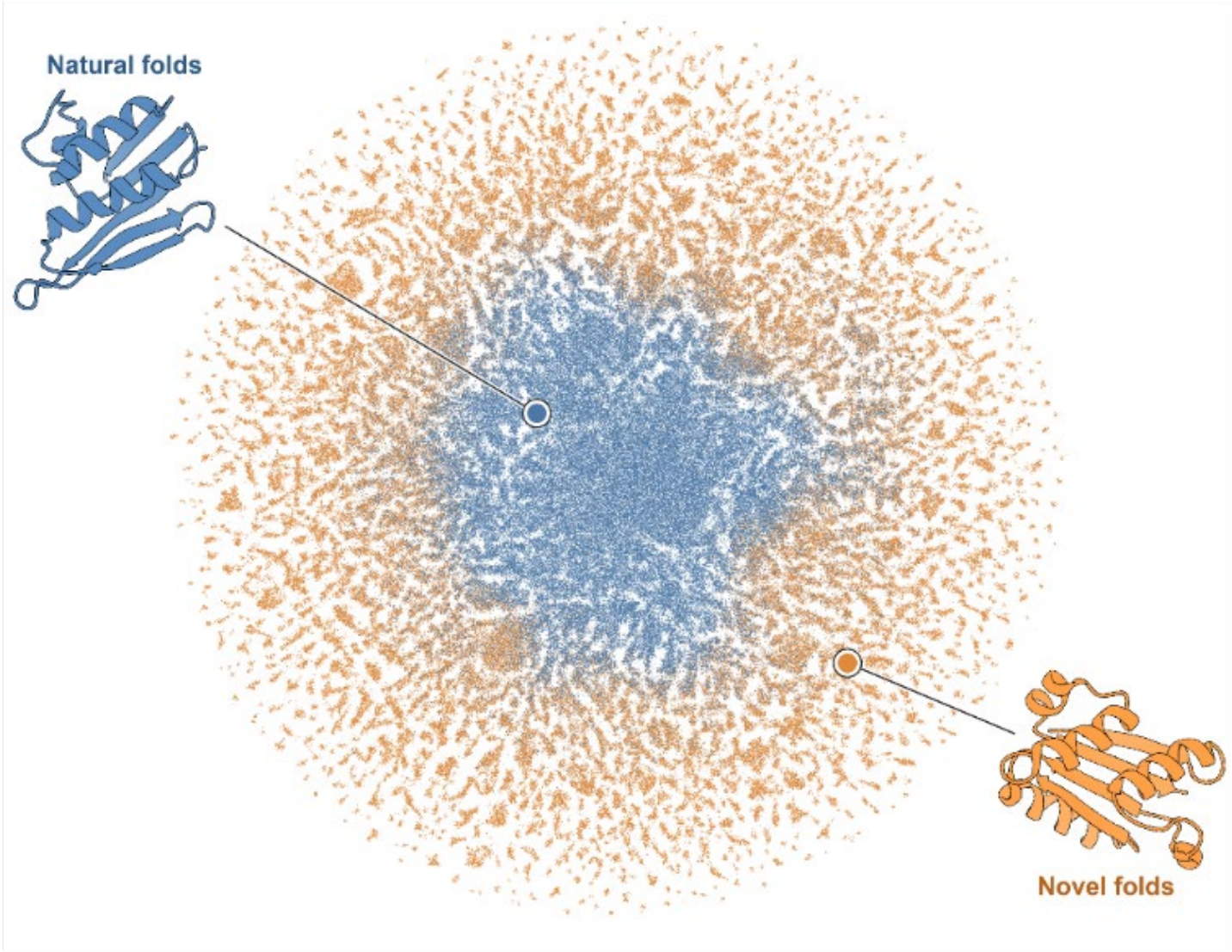

**Figure 1. The protein fold space**.

In the schematic representation, each dot represents a protein. Protein folds found in nature occupy only a portion of this space (blue dots), and numerous other stable folds are possible (orange).

## 2. Proteins evolved from random sequences often adopt novel folds

When evolution of globular proteins from random sequences was simulated from first principles, approximately half of the evolved proteins did not match any known or modeled structures as shown by searching PDB and AlphaFold DataBase (AFDB) using Foldseek (Sahakyan *et al.*, 2025). These observations imply that, when a new protein evolves *de novo* under selective pressure for foldability, which was presumably the main driving factor in early protein evolution (Makarov *et al.*, 2023) and accounts for most of the evolutionary conservation in protein families (Dokholyan & Shakhnovich, 2001), there is a high likelihood that it will have a fold not observed in nature. Calculating the exact probability of novel fold emergence when sampling random sequences using current structure prediction methods is complicated by two opposing factors. On one hand, the true fraction of novel folds could be lower than estimated because structure prediction methods are imperfect, and some structures might not actually be foldable, even if predicted with high confidence. On the other hand, the true fraction might be higher because the structure prediction models are trained on existing protein structures and tend to perform better for proteins resembling natural folds, whereas many proteins with novel folds might not be predicted with high confidence even if they are foldable and stable (Kinch *et al.*, 2021).

Another, and more fundamental problem is that there is no universal metric for accurately determining whether a protein has a novel fold. Moreover, protein fold is, to some extent, a subjective category, in contrast to a superfamily, which implies common origin and a specific function. An oversimplified solution is to require no hits to natural proteins in structural searches, even at a low similarity threshold; however, this criterion yields many false negatives because some low-similarity hits actually have the same fold architectures. For example, many proteins with the same fold have TM-score below 0.5, which is commonly accepted as a similarity threshold for proteins with the same folds (Xu & Zhang, 2010), while proteins with different folds may have TM-score above 0.5 (Edgar & Sahakyan, 2025). This is an inherently difficult problem because, even though the protein fold space cannot be genuinely continuous given the finite number of possible sequences, multiple intermediate states might exist between two distinct folds that gradually transition from one fold to the other. Selecting an optimal similarity cutoff that balances false positives against false negatives is therefore nontrivial.

These caveats notwithstanding, the *in silico* evolved proteins with novel folds were found to be as foldable and stable as those with folds found in nature (Sahakyan *et al.*, 2025). Furthermore, the number of mutations that had to be introduced into the initially random, evolving sequences did not appreciably differ between the known and novel folds. While artificial proteins with known folds required, on average, 407 mutations to evolve, proteins with novel folds required, on average, 389 mutations. Moreover, for computationally designed novel folds, the success rate of experimental validation was the same as for structures with known folds (Minami *et al.*, 2023). These observations imply that the apparent absence (or extreme rarity) of those novel folds in living organisms is not caused by their biophysical inferiority to the known folds or by the difficulty of reaching them in the protein structure space.

## 3. *De novo* emergence of proteins leads to novel folds

Understanding the routes of protein fold evolution is crucial for explaining why some fold topologies are common across all three domains of life, others are specific to narrow taxonomic lineages, and yet others have not been observed in nature at all. Exaptation of existing folds, whereby protein sequences and structures diverge until the ancestral fold becomes unrecognizable, is a common scenario for the emergence of protein folds qualifying as new. Sometimes, the changes are gradual and can be followed by sequence comparison and phylogenetic analyses; in other cases, the changes are more dramatic and abrupt, and can be traced only by structure comparison methods (Grishin, 2001). In yet other cases, the sequence changes are small but can trigger dramatic structural transformation leading to fold switching (Chakravarty *et al.*, 2026). The common mechanism in all these cases is exaptation of existing proteins.

A fundamentally different mechanism involves *de novo* genes that emerge from previously noncoding sequences, often in intergenic regions, and encode small proteins that can become functional and subject to purifying selection under certain conditions (Fesenko *et al.*, 2025, Tassios *et al.*, 2026, Aubel *et al.*, 2024). The arms race during host-pathogen co-evolution, particularly between bacteriophages and bacteria, is a common situation in which new proteins appear to emerge frequently (Nomburg *et al.*, 2024). Under strong selective pressure, even a small advantage in attack or defense can be sufficient for a newborn gene encoding a new protein to be fixed and propagated. Most likely, many rare protein folds originate by *de novo* gene birth rather than through divergence and repurposing of existing folds.

Anti-CRISPR (Acr) proteins that are encoded by diverse bacterial and archaeal viruses, and inactivate CRISPR-Cas adaptive immunity systems (Davidson *et al.*, 2020, Pawluk *et al.*, 2018) often show no structural similarity to any known folds (Sahakyan *et al.*, 2023). Experimentally solved structures provide support for the extensive structural novelty of Acrs, which inhibit several CRISPR–Cas types. In particular, NMR structures of AcrIF1 and AcrIF7 revealed compact folds without significant structural matches (Maxwell *et al.*, 2016, Kim *et al.*, 2020). Crystallography yielded additional examples, including AcrIF13, AcrIF23, and AcrIIA18, whose unusual architectures were not similar to those of any known proteins (Wang, Gao, *et al.*, 2022, Ren *et al.*, 2022, Wang, Li, *et al.*, 2022). Cryo-electron microscopy expanded this list of apparently unique folds to include AcrIF4, AcrIIC5, AcrVIA1, and AcrIIA26 (Gabel *et al.*, 2021, Sun *et al.*, 2023, Meeske *et al.*, 2020, Zheng *et al.*, 2026), which inhibit distinct CRISPR effectors. Although structure comparison analyses revealed possible homologs for some of the Acrs, including distinct connections with toxin-antitoxin systems, many of these viral proteins still lack homologs, and their origin remains unknown (Davidson *et al.*, 2020). A systematic search with sequence profiles combined with structure similarity search with computationally predicted models for 102 Acrs families discovered only 13 potential homologs, in addition to 14 Acrs for which putative homologs were known, whereas 75 of these Acr families did not match any structures. These observations suggest that Acrs with different novel folds emerged independently, on multiple occasions, in different viral taxa (Sahakyan *et al.*, 2023).

There are additional cases of apparent origin of novel protein folds *de novo*, in particular, by overprinting, that is, utilization of alternative open reading frames in pre-existing genes, which is relatively common in small viruses, such as microviruses (Rancurel *et al.*, 2009).

**4. The likelihood of a novel fold emerging increases with protein size**

If protein folds were randomly sampled from the structural space, larger proteins would be expected to adopt novel folds more often than smaller ones because more secondary structure elements have access to a larger combinatorial space and hence a greater number of possible topologies (Fig. 2). These considerations apply in cases when protein homology cannot be inferred from sequence similarity searches, but the proteins of interest nevertheless have similar structures. Such situations result from either divergent evolution leading to unrecognizable sequence similarity while the fold is preserved, or from convergent evolution leading two independently evolving proteins to the same fold. For small protein domains, such as helix-turn-helix (HTH), WW, or SH3, convergence is likely to be common as illustrated by the study of protein evolution from random sequences discussed above, whereas convergent evolution of larger folds is much less likely (Fig. 2). An important exception are proteins with repeating patterns and low complexity, such as β-propellers and α-solenoids, which evolve through multiple duplications of small structural elements (Pereira & Lupas, 2022, Fournier *et al.*, 2013). The minimal repeating unit of these folds, for example, a single β-sheet (blade) in β-propellers or an α-hairpin in α-solenoids, can emerge independently on multiple occasions and evolve into similar higher-order structures after a series of duplications (Smock *et al.*, 2016). In principle, a probabilistic model could be developed to estimate the likelihood of convergent evolution between two protein folds based on their size and complexity, but the algorithms for achieving this goal remain to be developed.

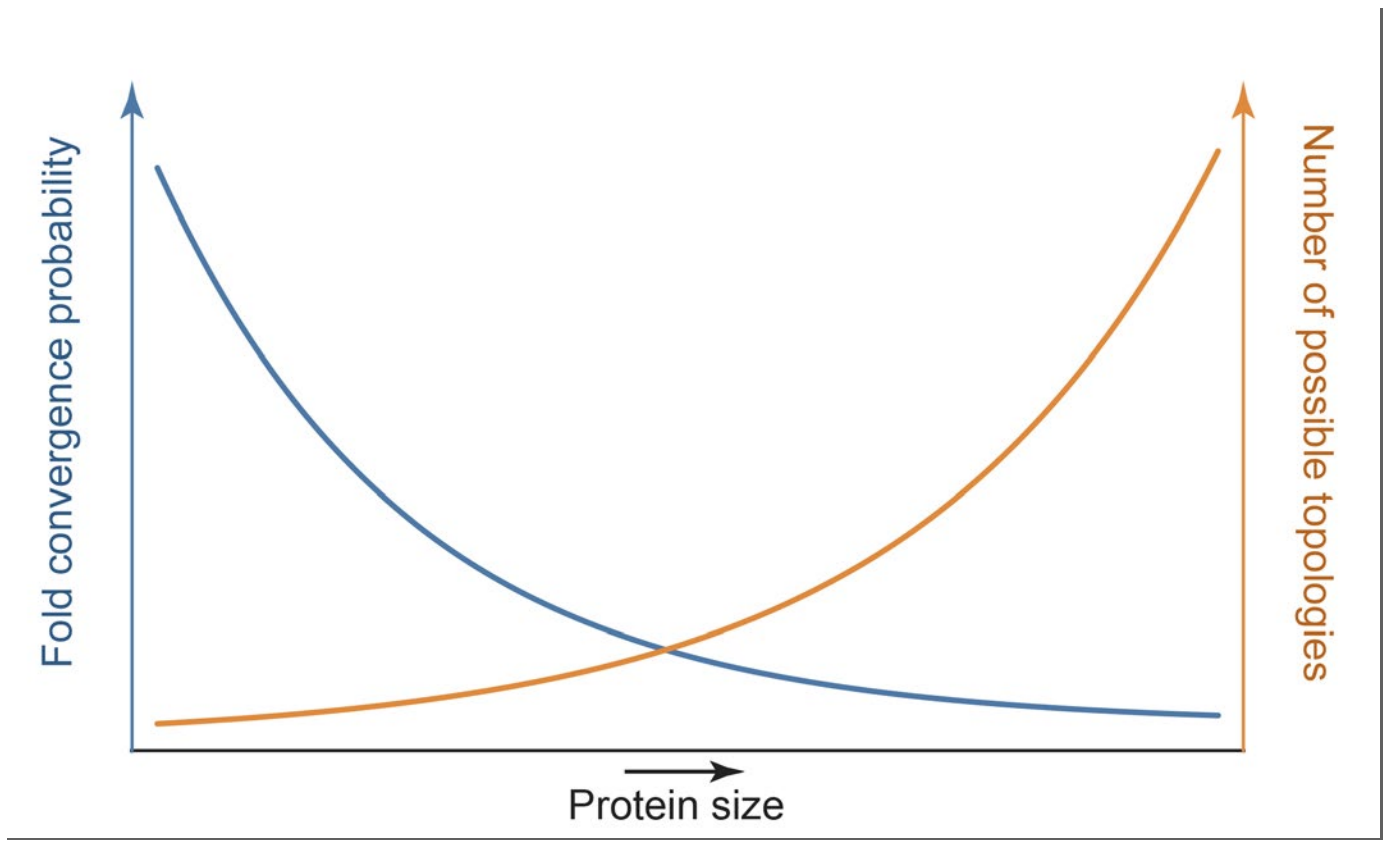

**Figure 2. Convergence in protein fold evolution depending on protein size.**

The probability of convergent evolution of convergent evolution of the same protein fold as a function of protein size is illustrated schematically. A greater number of secondary structure elements provides more possible combinations for forming a protein fold. As the number of possible folds increases, the probability of convergent evolution to the same fold drops.

## 5. Are novel folds useful?

We presented arguments supporting the notion that the space of all possible protein folds is much – apparently, vastly – larger than the limited set of folds observed in nature and that evolution has not explored a substantial fraction of that structural space. A fascinating question, however, is whether these novel folds are biologically and/or biotechnologically useful. The immediate answer from evolution's appears to be that they are not: it seems to be easier to graft a new function onto an existing fold, which is why the same fold often performs different, sometimes, completely unrelated, functions. However, as discussed above, novel folds do not appear to be intrinsically 'worse' than those existing in nature, and therefore, in certain cases, could be more efficient than existing ones. Given the need for new solutions to critical problems, whether it is new enzymes for degrading plastic, new biosensors, or therapeutics, novel folds might offer new or better solutions. There are already several examples of designed functional proteins with novel folds. Castro et al. designed a small protein with a novel fold that contains three distinct, nonoverlapping motifs from respiratory syncytial virus fusion protein epitopes and interacts with three different antibodies (Castro *et al.*, 2025). Such a design of chimeric multiepitope proteins has the potential to improve cocktail vaccine strategies. Other examples include active enzymes such as a serine hydrolase and a protease with a novel fold designed with molecular modeling methods (Lauko *et al.*, 2025, Choi *et al.*, 2026). The catalytic sites of these enzymes were used to design a backbone that maintained the key residues of the active site in the correct configuration necessary for enzymatic activity. These results show that the active catalytic sites are not necessarily coupled to their native folds, and that novel folds can host them, offering modified and/or improved properties.

The fact that many proteins generated by protein design tools have novel folds suggests that these models learn general principles of protein structural organization rather than merely memorizing existing protein structures. However, even for state-of-the-art structure prediction models, accuracy remains limited: they predict correct structures for some *de novo* folds but not the majority (Buchel *et al.*, 2026). The accuracy largely depends on the availability of evolutionary information, which is unavailable for novel folds except in *in silico* or *in vitro* evolutionary experiments. Exploration of the uncharted regions of the protein fold space can be expected to provide for training more generalizable models for protein structure prediction and design. Thus, to understand the protein sequence-structure relationship in general, beyond what has been shaped by natural evolution, we need to explore artificial proteins that are biophysically plausible but do not seem to exist in nature.

## 6. Conclusions

Why are there so few protein folds in nature compared to the vast number of possible ones, many of which appear not to be inferior to those found in living organisms by any available metric? We present here several lines of evidence suggesting that the reason is evolutionary contingency: it is, in general, advantageous for evolution to tinker with already evolved folds rather than to invent new ones. Nevertheless, new folds do emerge during evolution, especially, in the context of host-parasite arms races. Furthermore, exploration of the regions of the protein structural space that are not populated by existing folds or are only inhabited by rare folds can help both to better understand principles of protein fold evolution and to design better, more general models for prediction, design and analysis of protein structures.

**Funding information**

This work was supported by the Intramural Research Program of the NIH of the USA (National Library of Medicine).

**Competing interests**

The authors declare no competing interest.